\documentclass[9pt]{article}
\usepackage{spconf,amsmath,graphicx, xcolor}
\usepackage{algorithm,algpseudocode}
\usepackage{amsmath}
\usepackage{amsfonts}
\usepackage{amssymb}
\usepackage{amsmath,amsfonts,amssymb,amsthm}
\usepackage{bbm}
\usepackage{subcaption}

\algnewcommand{\IIf}[1]{\State\algorithmicif\ #1}
\algnewcommand{\EndIIf}{\unskip\ \algorithmicend\ \algorithmicif}
\def\real{\mathbb{R}}

\def\X{\mathbf{X}}

\def\P{\mathbf{P}}

\def\Y{\mathbf{Y}}

\def\B{\mathbf{B}}
\def\I{\mathbf{I}}
\def\y{\mathbf{y}}
\def\x{\mathbf{x}}
\def\W{\mathbf{W}}

\usepackage{diagbox}
\DeclareMathOperator*{\argmin}{\mathrm{argmin}\,\,}
\newenvironment{psmallmatrix}
{\left(\begin{smallmatrix}}
{\end{smallmatrix}\right)}

\newtheorem{proposition}{Proposition}
\begin{document}
\title{\lowercase{$r$}-local Unlabeled Sensing: Improved algorithm and applications }
\name{Ahmed Ali Abbasi$^{ \dagger}$ \qquad Abiy Tasissa$^{\star}$ \qquad Shuchin Aeron$^{\dagger}$} 
\address{$^{\dagger}$ Department of ECE, Tufts University,
Medford, MA 02155 \\
$^{\star}$ Department of Mathematics, Tufts University, Medford, MA 02155 \\
}
\maketitle
\begin{abstract}
  The unlabeled sensing problem is to solve a noisy linear system of equations under unknown permutation of the measurements. We study a particular case of the problem where the permutations are restricted to be $r$-local, i.e. the permutation matrix is block diagonal with $r\times r$ blocks. Assuming a Gaussian measurement matrix, we argue that the $r$-local permutation model is more challenging compared to a recent sparse permutation model.   We propose a proximal alternating minimization algorithm for the general unlabeled sensing problem that provably converges to a first order stationary point. Applied to the $r$-local model, we show that the resulting algorithm is efficient.  We validate the algorithm on synthetic and real datasets. We also formulate the $1$-d unassigned distance geometry problem as an unlabeled sensing problem with a structured measurement matrix.
\end{abstract}
\vspace{-0.5em}
\section{Introduction}
\vspace{-0.5em}
\label{sec:intro}
The standard least squares problem is to recover signal  $\X^* \in \real^{d \times m}$ given measurements $\Y \in \real^{n \times m}$  and the measurement matrix $\B$. The signal matrix $\X^*$ is estimated by solving the least squares minimization problem $\min_\X\,  \lVert \Y- \B\X \rVert_{F}^2$. This formulation assumes that there is one to one correspondence between measurements (the rows of $\Y$) and the measurement vectors (the rows of $\B$), i.e., we know which measurement corresponds to what measurement vector. 
However, in many problems of applied interest such as header free communication for mobile wireless networks \cite{header_free},  sampling in the presence of clock jitter \cite{Balakrishnan},  linear regression without correspondence \cite{broken} and point cloud registration \cite{pose_and_correspondence}, this mapping is not explicitly given. This motivates the study of the unlabeled sensing problem with the prototypical form given by 
\begin{equation}
\Y = \P^* \B \X^* + \W, \label{eq:model}
\end{equation}
where $\P \in \real^{n\times n}$ denotes the unknown permutation matrix and $\W\in \real^{n\times m}$ denotes an additive Gaussian noise with $\W_{ij} = \mathcal{N}(0,\sigma^2)$.  
The problem is to estimate $\X^*$ and $\P^*$ given the measurement matrix $\B$ and measurements $\Y$.
\footnote{ \tiny© 2022 IEEE.  Personal use of this material is permitted.  Permission from IEEE must be obtained for all other uses, in any current or future media, including reprinting/republishing this material for advertising or promotional purposes, creating new collective works, for resale or redistribution to servers or lists, or reuse of any copyrighted component of this work in other works.}

\textbf{Related work}:
The unlabeled sensing problem was first considered with a Gaussian measurement matrix $\B$ in the single-view setup, where $m=1$, in \cite{unnikrishnan2018unlabeled}. There, the authors showed that $n=2d$ noiseless measurements are necessary and sufficient for recovery of $\x^*$. The same result was subsequently proven in \cite{Dokmanic} using an algebraic argument. For the case of Gaussian $\mathbf{B}$, the work in \cite{pananjady} showed that with $\textrm{SNR} \triangleq  \lVert \X^* \rVert_F^2/m\sigma^2$, $\textrm{SNR} = \Omega(n^2)$ is necessary for recovering $\P^*$ exactly.  For the multi-view setup, where $m>1$, the result in \cite{snr} shows that the necessary $\textrm{SNR}$ for recovery decreases as $m$ increases, with necessary $\textrm{SNR} = O(1)$  for $m = \Omega(\log n)$. 

\begin{figure}[t!]
    \centering
    \hspace{-4.5cm}\begin{subfigure}{0.16\textwidth}
    \begin{center}
    {\includegraphics[width = 0.95 \linewidth]{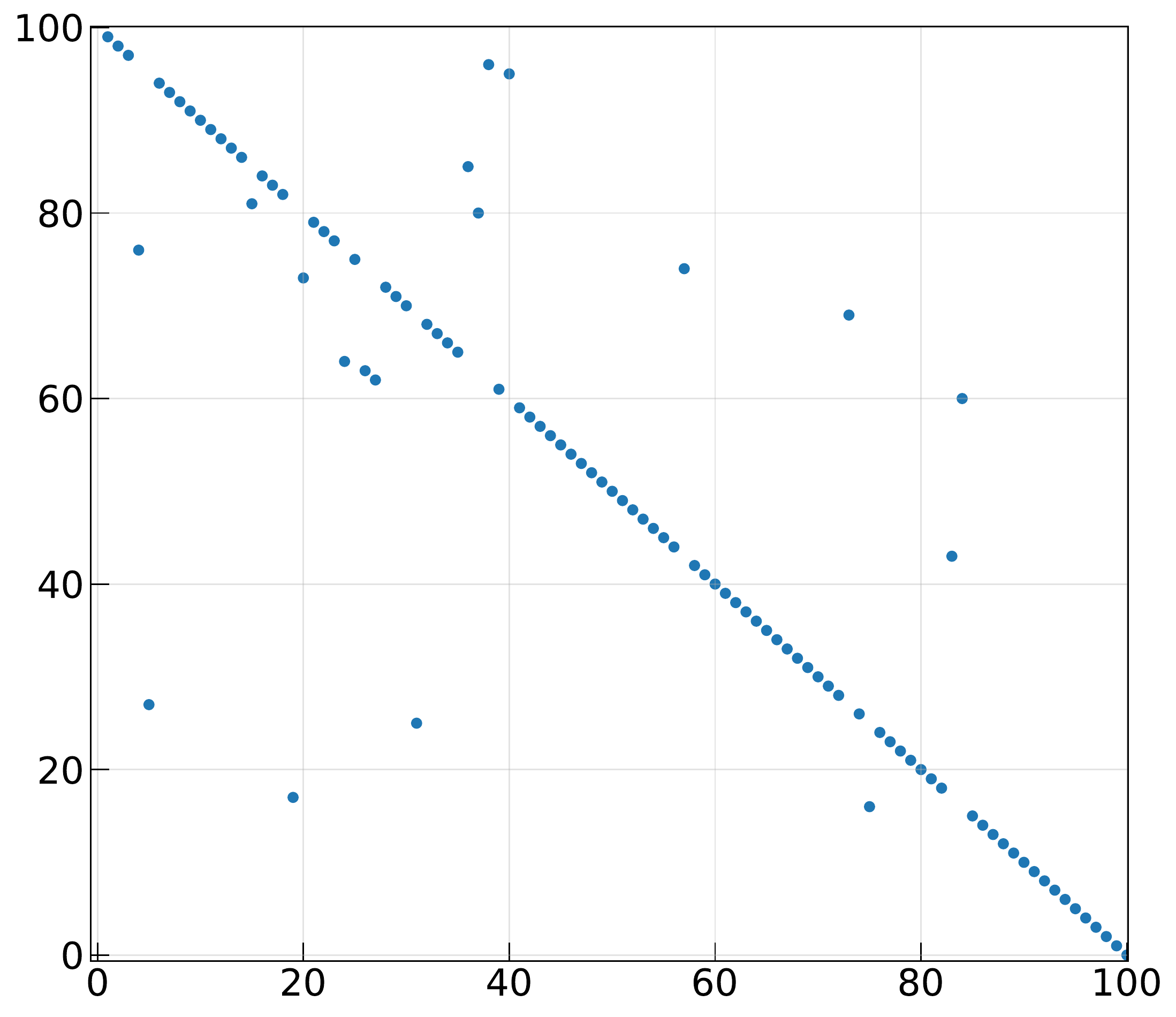}}
    \caption{}
    \label{fig:k_sparse}
    \end{center}
    \end{subfigure}\linebreak
\begin{subfigure}{0.16\textwidth}
    \begin{center}
    {\includegraphics[width = 0.95 \linewidth]{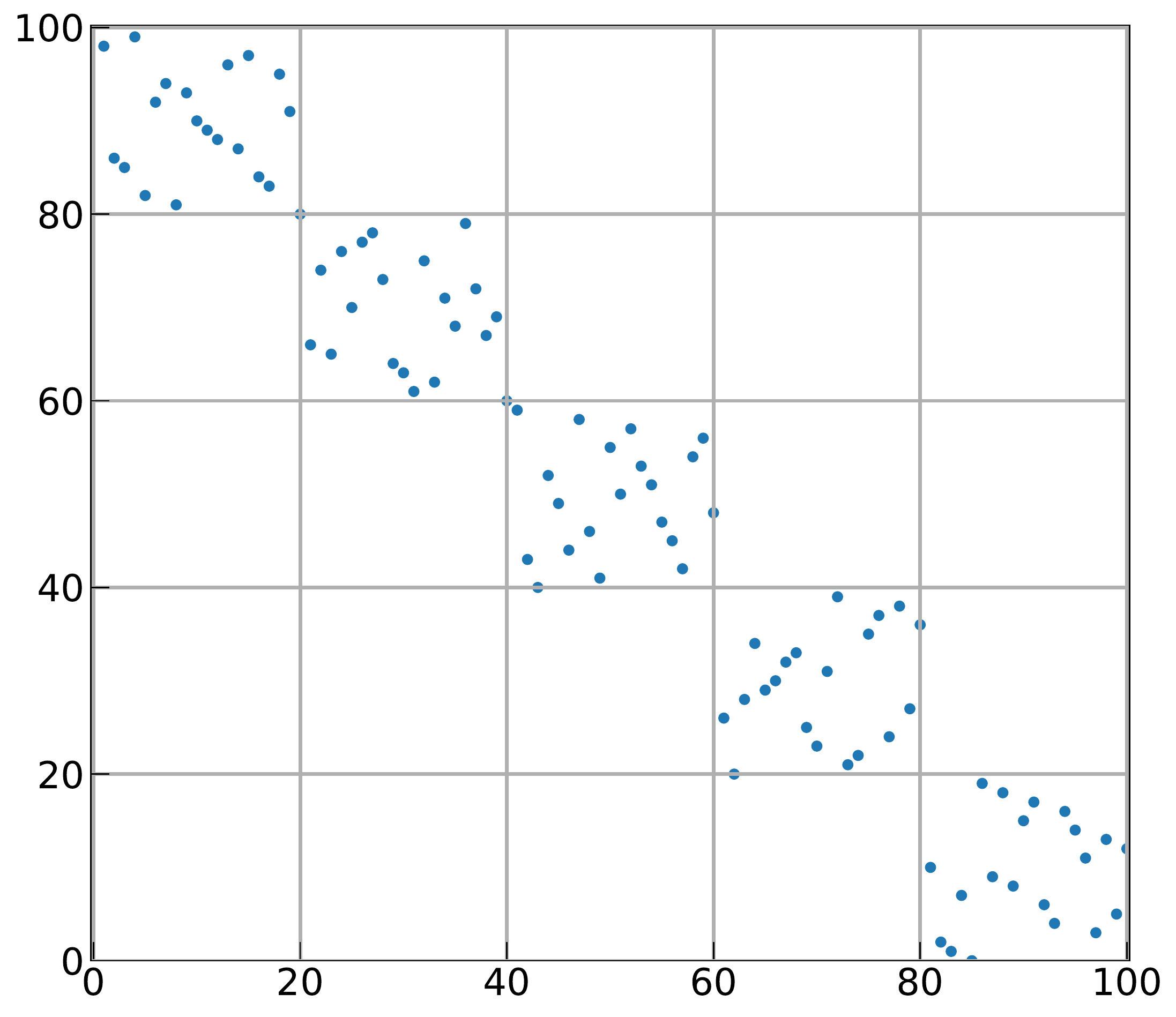}}
    \caption{}
    \label{fig:r_local}
    \end{center}
    \end{subfigure}   
    \begin{subfigure}{0.25\textwidth}
    \begin{center}\vspace{-2.45cm}
    {\includegraphics[width = 0.99 \linewidth]{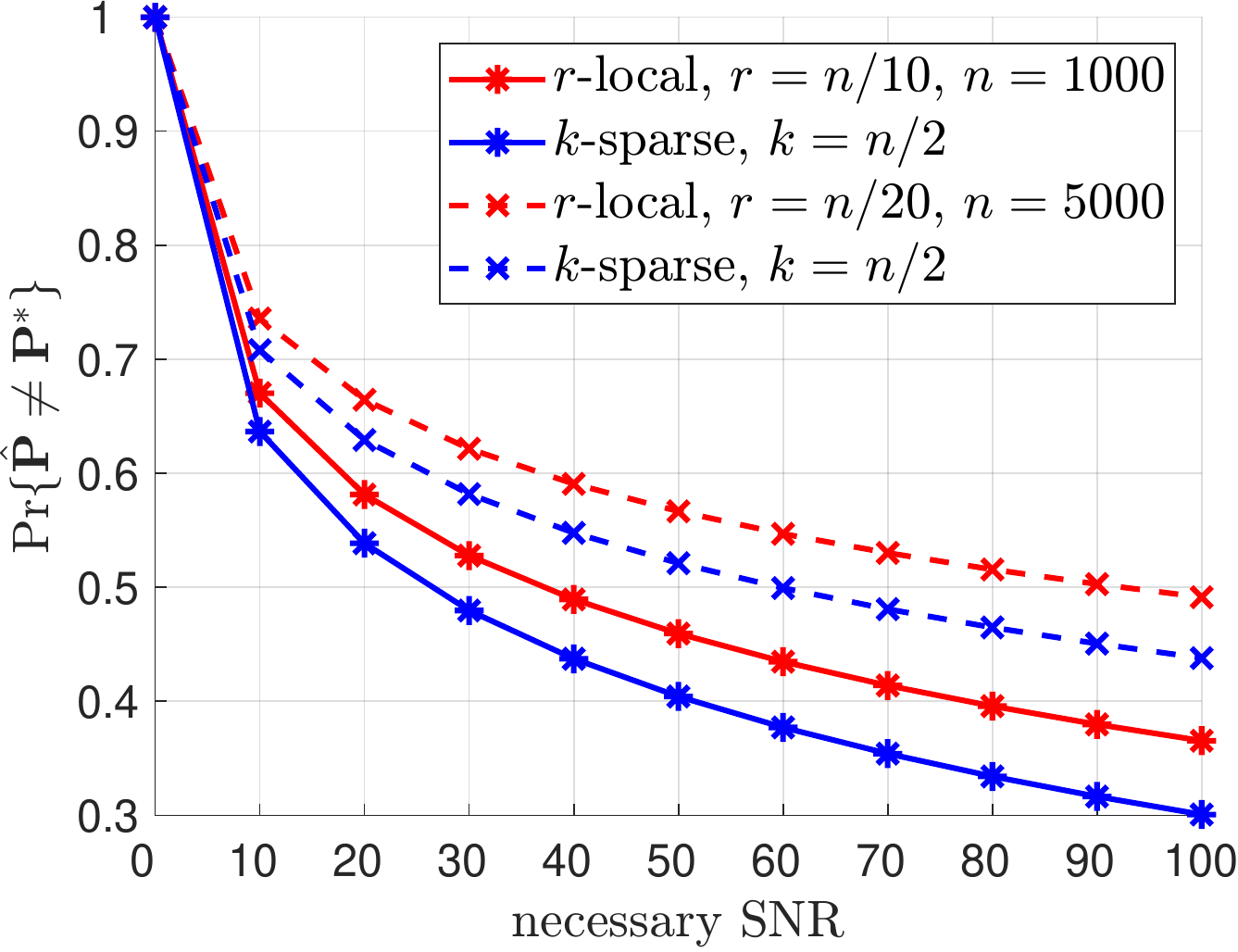}}
    \caption{}
    \label{fig:entropy}
    \label{entropy}
    \end{center}
    \end{subfigure}   
    \caption{(a). A sparse (or partially shuffled) permutation considered in \cite{snr,slawski_two_stage,zhang2019permutation,slawski2020sparse}. (b). The $r$-local permutation structure considered in this paper, with $r=20$. (c). Figure plots the lower bound on the event $\Pr\{\widehat{\P} \neq \P^* \}$ against $\textrm{SNR}$ for the two models (see equation \eqref{eq:inach}). }
    \label{fig:models}
    \vspace{-1.25em}
\end{figure}

Several algorithms have been proposed for the single view  unlabeled sensing problem \cite{header_free,pananjady,concave}.  Algorithms for the  multi-view problem setup have been proposed in works \cite{snr,slawski_two_stage,Levsort,zhang2019permutation,ojsp}. As computing the maximum likelihood estimate of $\P^*$ may be NP-hard \cite{Levsort}, the works in \cite{snr,slawski_two_stage,zhang2019permutation} impose a $k$-sparse structure on $\P^*$ (see Figure \ref{fig:k_sparse}) where $n-k$ elements of the permutation are on the diagonal. In  \cite{ojsp}, the authors introduced the $r$-local model (see section \ref{sec:r_local_model}) along with an algorithm that is based on graph matching. A permutation matrix $\P^*$ of size $n \times n$ is $r$-local if it is composed of $n/r$ blocks. Formally, $\P^* = \textrm{diag}\left(\P^*_1, \cdots, \P^*_{n/r}\right)$, where $\P^*_{i} \in \Pi_r$ denotes an $r \times r$ permutation matrix (see Figure \ref{fig:r_local}). 

\textbf{Applications}:  In this paper, we  consider two applications of the $r$-local unlabeled sensing model. The first application is the unassigned distance geometry problem (uDGP) \cite{Huang_2021} (see section \ref{sec:udgp}). The second application is image unscrambling \cite{gumbel_sinkhorn} (see section \ref{sec:results}). Another application of the $r$-local permutation model is linear regression with blocking \cite{lahiri2005regression}.

\textbf{Contributions}:  (a) We compare the $r$-local model with the $k$-sparse model and argue that the $r$-local problem is more challenging under the most widely studied case of Gaussian measurement matrix.
(b) We formulate the $1$-d unassigned distance geometry problem (uDGP) \cite{Huang_2021} as an unlabeled sensing problem with a structured measurement matrix.  (c) We propose a proximal alternating minimization algorithm for the $r$-local unlabeled sensing problem. Simulation results show that the proposed algorithm outperforms the algorithms in \cite{snr,Levsort,ojsp,zhang2019permutation}. 
\vspace{-0.65em}
\section{$\lowercase{r}$-local model vs $\lowercase{k}$-sparse model}
\label{sec:r_local_model}
\vspace{-0.65em}
A permutation matrix $\P$ is $k$-sparse if $n-k$ of its elements are on the diagonal, i.e., $\langle \mathbf{I},\P \rangle  = n - k$. Figures \ref{fig:k_sparse} and \ref{fig:r_local}  show a $k$-sparse and an $r$-local permutation matrix, respectively. We compare the two models by the difficulty of recovering $\P^*$ under each model when the matrix $\mathbf{B}$ is assumed to be random Gaussian. Under this assumption, for any estimator $\widehat{\P}$, the probability of error $\Pr\{\widehat{\P} \neq \P^* \}$ is lower bounded by the result in \cite{snr}
\begin{equation}
 \Pr\{\widehat{\P} \neq \P^*\} \geq 1 - \frac{1 + \frac{nm}{2}\, \log_2(1 + \textrm{SNR})}{H(\P^*)},
 \label{eq:inach}
\end{equation}
where $\P^*$ is a random variable and $H(\P^*)$ denotes the entropy of $\P^*$. For uniformly distributed $\P^*$, the entropy is $H(\P^*)= \log_2 a$, where $a$ is the  number of possible permutations.  Without any assumption on the permutation structure, the entropy of $\P^* \in \mathbb{R}^{n \times n}$ is $H(\P^*) = \log_2(n!)$. The entropy for $r$-local permutations is $H_r(\P^*) = \log_2 r!^{n/r}$. The entropy for $k$-sparse permutations is $H_k(\P^*) = \log_2(n!/(n-k)!)$. Performance guarantees for proposed algorithms are given  for $k={n}/{8}$ in \cite{snr} and {$k={n}/{4}$} in \cite{zhang2019permutation}. All simulation results for permutation recovery in \cite{slawski_two_stage,slawski2020sparse} are for $k
\le n/2$.  Figure \ref{fig:entropy} shows that, for the same $\textrm{SNR}$, the lower bound in \eqref{eq:inach} is higher for the $r$-local model,  $r=\{n/10,n/20\}$, than for the $k$-sparse model, with $k=n/2$. This suggests that permutation recovery under the $r$-local model is challenging in the large $n$ regime. 
\vspace{-1em}\section{1-D Unassigned Distance Geometry as an unlabeled sensing problem}
\label{sec:udgp}
\begin{figure}[!]
    \begin{center}
    {\includegraphics[width = 0.95 \linewidth]{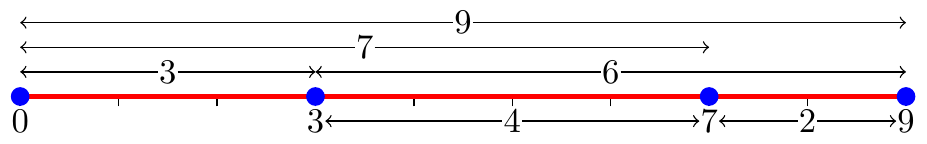}}
    \caption{The $1$-d unassigned distance geometry problem (uDGP) is to recover the point coordinates $(0,3,7,9)$ from their unlabeled pairwise distances $\{3,2,4,6,7,9\}$. uDGP can be formulated as an unlabeled sensing problem with a deterministic measurement matrix $\B_u$, see equation \eqref{eq:B_u}.} 
    \label{fig:udgp_eg}
    \end{center}
    \vspace{-1.05em}
\end{figure}

 The one-dimensional unassigned distance geometry problem (1-d uDGP) (see Figure \ref{fig:udgp_eg}) is to recover point coordinates from their unlabeled pairwise distances \cite{Huang_2021,duxbury2021unassigned,dakic2000turnpike}. The distances are unlabeled as the pair $(i,j)$ corresponding to the distance $\lvert x_i - x_j \rvert$ is not known. Let $\overline{\x} \in \real^{d} = [\overline{x}_1, \cdots,\overline{x}_d]^{\intercal}$ be the vector of unknown coordinates. Without loss of generality, we assume that the unknown point coordinates are sorted in decreasing order. It follows that each pairwise distance $\overline{x}_i - \overline{x}_j$ is a linear measurement of $\overline{\x}$. As the distances are translation invariant, we can also assume that the left most point is at the origin. The distance vector $\y \in \real^{d(d-1)/2}$ is the matrix-vector product $\y = \B_u\x^*$ where $\x^*  \in \real^{d-1}$. It can be verified that $\B_u$ has the following form. 
\begin{align}
&\B_u = [\mathbf{I}_{(d - 1) \times (d-1)};\B_2;\cdots; \B_{d-1}], \nonumber\\
& [\mathbf{B}_{i} \in \real^{(d-i) \times (d-1)}]_{pq} = \begin{cases}
    1  &\text{if $q = i-1$} \\
    -1 & \text{if $q = p + i-1$}\\
    0 & \text{otherwise},
\end{cases}
\label{eq:B_u}
\end{align}
where $[\mathbf{A}_1;\mathbf{A}_2]$ denotes vertical concatenation of the matrices $\mathbf{A}_1,\mathbf{A}_2$. The unlabeled sensing formulation for the example in Figure \ref{fig:udgp_eg}, with $d=4$ points, is given below\vspace{-0.465em}
$$
\underset{\y}{\begin{psmallmatrix}
3\\
2\\
4\\
6\\
7\\
9\\
\end{psmallmatrix}} = 
\underset{\P^*}{
\begin{psmallmatrix}
0 & 0 & 1 & 0 & 0 & 0\\
0 & 0 & 0 & 1 & 0 & 0\\
0 & 0 & 0 & 0 & 0 & 1\\
0 & 0 & 0 & 0 & 1 & 0\\
0 & 1 & 0 & 0 & 0 & 0\\
1 & 0 & 0 & 0 & 0 & 0\\
\end{psmallmatrix}
}
\underset{\B_u}{
\begin{psmallmatrix}
1 & 0 & 0 \\
0 & 1 & 0 \\
0 & 0 & 1 \\
1 & -1 &0\\
1 &  0 &-1 \\
0 &  1 &-1 \\
\end{psmallmatrix}}
\underset{\x^*}{
\begin{psmallmatrix}
9\\
7\\
3\\
\end{psmallmatrix}.
}
$$
\textbf{Why $r$-local model for uDGP?} Imposing an $r$-local structure on $\P^*$ has a practical interpretation for the uDGP problem. The problem where, in addition to the distances $\lvert x_i - x_j \rvert$, one of the two corresponding indices $(i,j)$ is known, is modelled by an $r$-local $\P^* = \textrm{blkdiag}(d-1,\cdots,2$). The $r$-local structure also models the problem where distance assignments are known up to a cluster of points. For example, for two clusters, each distance is known to be between a pair of points in cluster $1$ or cluster $2$ (intra-cluster) or a point in cluster $1$ and a point in cluster $2$ (inter-cluster) but the  pair of points corresponding to each distance is still unknown.
\section{Proposed approach and algorithm }
\label{sec:algo}
In this section, we present a new algorithm for the $r$-local unlabeled sensing problem. To motivate our proposed algorithm, we note that for  i.i.d Gaussian noise $\W$ in \eqref{eq:model}, the maximum likelihood estimate of $\P^*$ is given by
 \begin{equation}
 \underset{\X,\P \in \Pi_n}{\arg\min}\, F(\X,\P) = \lVert \Y-\P \B\X\rVert_{F}^{2}, \label{eq:obj} 
 \end{equation}
 where $\Pi_n$ denotes the set of permutations of order $n$.  
 The optimization problem in \eqref{eq:obj} is non-convex since the set of permutations is discrete.  
 A natural optimization scheme for the problem in \eqref{eq:obj} is the alternating minimization algorithm which alternates between updating the signal matrix $\X$ and the permutation $\P$. The works in \cite{slawski_two_stage,Levsort,zhang2019permutation,slawski2020sparse} propose one step estimators for $\P^*,\X^*$.  We consider complete alternating minimization for \eqref{eq:obj} yielding the following updates
 \begin{align}
    &\P^{(t)}  = \argmin_{\P \in \Pi_n} \langle -\Y {\X^{(t)}}^{\intercal}\B^{\intercal},\P \rangle, \label{eq:P_update}\\
    &\X^{(t+1)} = \argmin_{\X}  F(\X,\P^{(t)}) = {(\P^{(t)}\B)}^{\dagger} \Y.
    \label{eq:x_update}
\end{align}
The iterates $(\P^{(t)},\X^{(t+1)})$ above are monotonically decreasing $F(\P^{(t+1)},\X^{(t+1)}) \leq F(\P^{(t)},\X^{(t+1)})$. However, this is not sufficient to establish convergence of the iterates. Specifically, we can not claim that ${\lim_{t \to \infty}} \{( \P^{(t)},\X^{(t)})\}$ exists. We  propose to use proximal alternating minimization (PAM) \cite{prox_alt_min} instead which, as we show in section \ref{subsec:convergence_analysis}, converges to a first order stationary point of the objective in \eqref{eq:obj}. PAM adds a regularization term that encourages the new iterate to be close to the current iterate. Formally, for the objective in \eqref{eq:obj}, the PAM updates are given by
\begin{align}
    &\P^{(t)} = \underset{\P \in \Pi_n}{\argmin}\,\, F(\X^{(t)},\P) + \lambda^{(t)} \lVert \P - \P^{(t-1)} \rVert_F^2 \label{eq:P_update_prox}, \\
   &\X^{(t+1)} =\underset{\X}{\argmin}\,\, F(\X,\P^{(t)}) + \lambda^{(t)} \lVert \X - \X^{(t)} \rVert_F^2, \label{eq:X_update_prox}
\end{align} 
where $\lambda^{(t)} > 0$. Our algorithm  follows from the updates in \eqref{eq:P_update_prox} and \eqref{eq:X_update_prox}. Similar to \eqref{eq:P_update} and \eqref{eq:x_update}, the updates are a linear program and a regularized least squares problem, respectively. Assuming an $r$-local structure on $\P^*$, the update in \eqref{eq:P_update_prox} is equivalent to the simpler update of $n/r$ blocks of size $r$. A summary of our algorithm is given in Algorithm \ref{algo:alt_min_prox}. 

\textbf{Algorithmic details}: We use the \emph{collapsed} initialization, as also done in \cite{ojsp}, to initialize $\widehat{\Y}^{(0)} = \B \widehat{\X}^{(0)}$. The initialization estimates $\widehat\X$ from $n/r$ measurements given by summing $r$ consecutive rows in $\B$ and $\Y$. The convergence criteria is based on the relative change in the objective  $(F^{(t-1)} - F^{(t)})/F^{(t-1)}$. The total cost of updating all the local permutations by linear programs is $O(nr^2)$. In line 8, $\dagger$ denotes the Moore-Penrose pseudo-inverse. The regularized least squares problem can be solved in $O(2nd^2-\frac{2}{3}d^3)$,    
\begin{algorithm}[t]
\caption{Proximal Alt-Min for $r$-local unlabeled sensing}
\label{algo:alt_min_prox}
\begin{algorithmic}[1]
\Require{$\B, \Y, \widehat{\Y}^{(0)}, \widehat{\X}^{(0)}, r, \lambda, \epsilon$}\ \vspace{0.215em}
\Ensure{$\widehat{\mathbf{\P}}$}\vspace{0.01em}
\State{$ t \gets 0$}
\State{$\widehat{\P}_k^{(t)} \gets \mathbf{1}_{r \times r}/r$}
\While {$\textrm{relative change} >  \epsilon $} 
    \For {$k \in 1 \cdots n/r$}\vspace{0.15em}
    \State{$\widehat{\P}_k = \underset{\P_k \in \Pi_r}{\argmin} - \langle  \Y \widehat{\Y}^{(t)} + \lambda \widehat{\P}^{(t)}_k,  \P_k   \rangle$}
    \EndFor
    \State{$\widehat{\mathbf{\P}}^{(t+1)} \gets  \text{diag} (\widehat{\P}_{1},\cdots,\widehat{\P}_{n/r})$} 
    \State{$\widehat{\X}^{(t+1)}  \gets {\big[\frac{\B_{n \times d}}{ \sqrt{\lambda} \I_{d \times d}}\big]}^{\dagger} \big[ \frac{\widehat{\P}^{(t+1)\intercal} \Y}{\sqrt{\lambda} \widehat{\mathbf{X}}^{(t)}}\big]$}
    \State{$\widehat{\Y}^{(t+1)}  \gets  \B \widehat{\X}^{(t+1)}$}
    \State{$t \gets t+1$}
    \EndWhile
\end{algorithmic}
\end{algorithm} 
\vspace{-1.5em}\subsection{Convergence analysis}
\label{subsec:convergence_analysis}
\vspace{-0.5em}\begin{proposition}
The sequence of iterates $\{(\X^{(t)},\P^{(t)})\}$ generated by \eqref{eq:P_update_prox}, \eqref{eq:X_update_prox} converges to a first order stationary point of the objective in \eqref{eq:obj}. 
\end{proposition}\vspace{-0.55em}
\textit{Proof.} The result follows from Theorem $9$ in \cite{prox_alt_min}.  To use Theorem $9$, we need to verify that the Kurdyka-Lojasiewicz (KL) inequality holds for the objective
$F(\X,\P) = \lVert \Y - \P \B \X \rVert_F^2 + \mathbf{I}_C(\P)$ where $\mathbf{I}_C(\P)$ is the indicator function of the set of permutations. First, note that the set of permutations of order $n$ is semi-algebraic because each of the $n!$ permutations can be specified via $n^2$ linear equality constraints. The indicator function of a semi-algebraic function is a semi-algebraic function \cite{prox_alt_min}. The first term in the objective is polynomial, and therefore real-analytic. The sum of  a real-analytic function and a semi-algebraic function is sub-analytic, and sub-analytic functions satisfy the KL inequality \cite{xu2013block}.  \vspace{-0.65em}
\section{Experiments}
\vspace{-0.65em}
\label{sec:results}
For all experiments using Algorithm \ref{algo:alt_min_prox}, we set the tolerance $\epsilon  = 0.01$. The proximal regularization is set to $\lambda^{(1)}=100$ and scaled as $\lambda^{(t+1)} = \lambda^{(t)}/10$.  MATLAB code for the experiments is available at the first author's github account: github.com/aabbas02/Proximal-Alt-Min-for-ULS-UDGP.
\begin{figure*}[t!]%
    \centering
    \begin{subfigure}[t]{0.24\textwidth}
    \begin{center}
    {\includegraphics[width = 0.99 \linewidth]{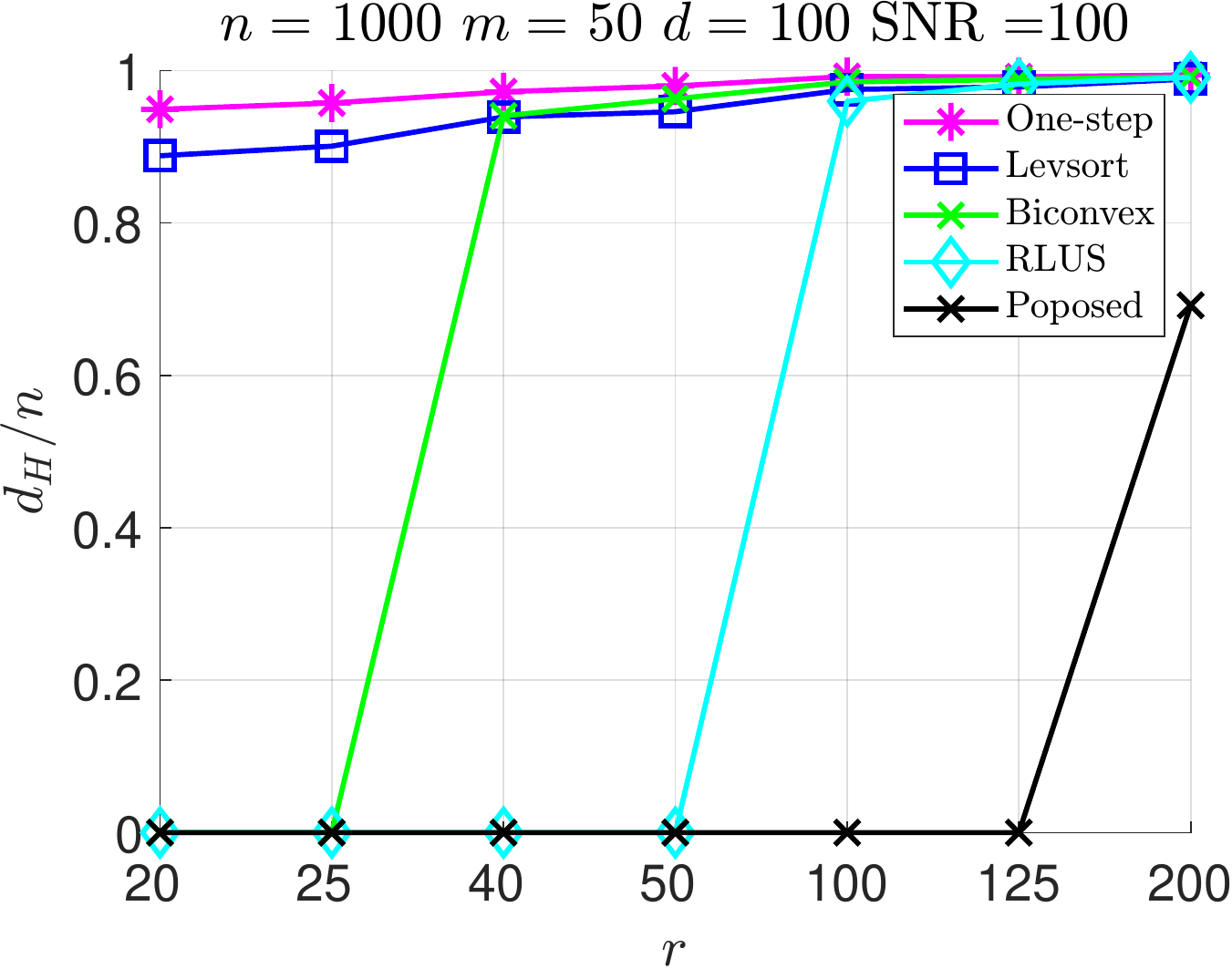}}
    \caption{}
    \label{fig:benchmark}
    \end{center}
    \end{subfigure}
    \begin{subfigure}[t]{0.23\textwidth}
    \begin{center}
    {\includegraphics[width = 0.99 \linewidth]{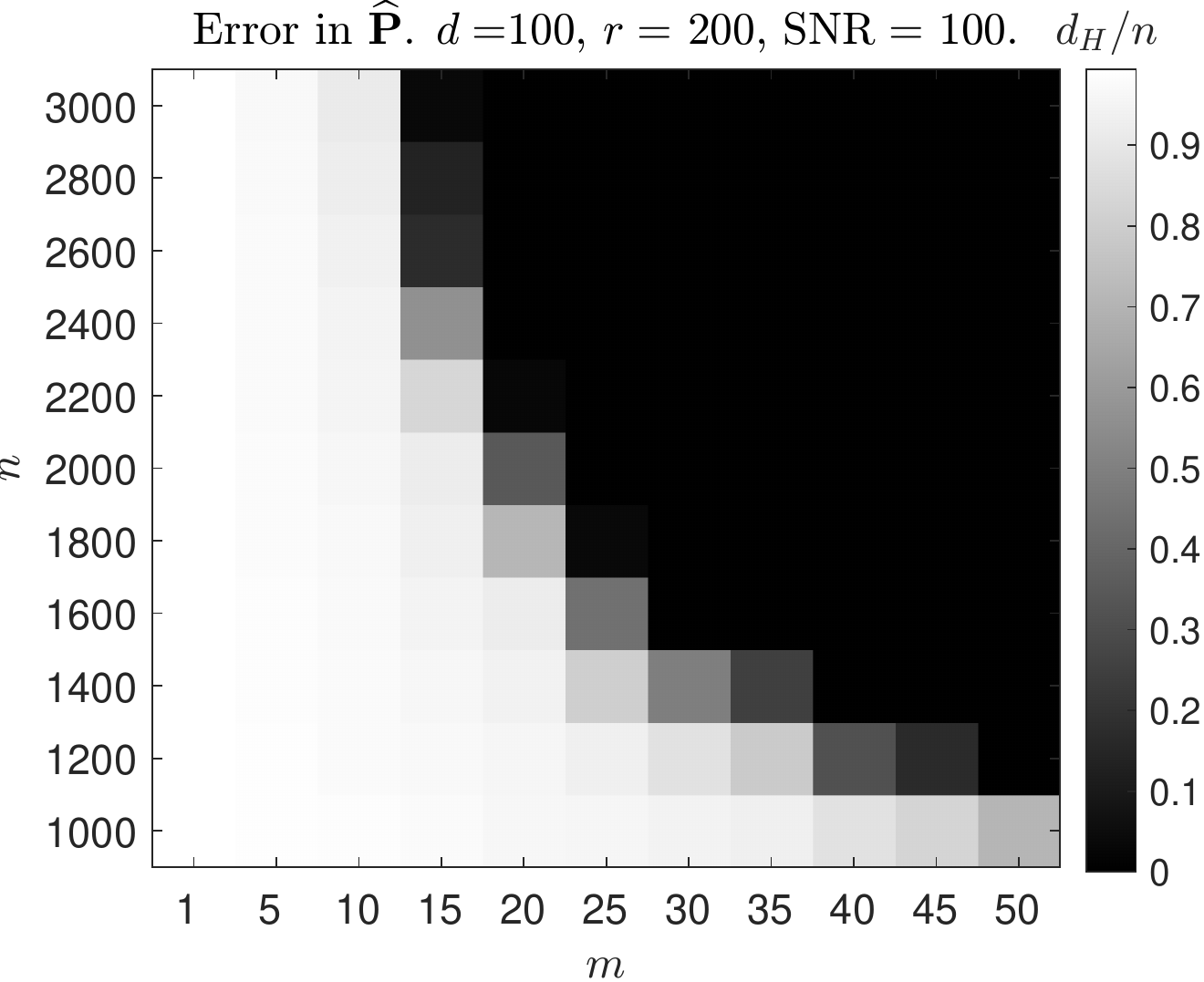}}
    \caption{}
    \label{fig:phase_transition_P}
    \end{center}
    \end{subfigure}
    \begin{subfigure}[t]{0.23\textwidth}
    \begin{center}
    {\includegraphics[width = 0.99 \linewidth]{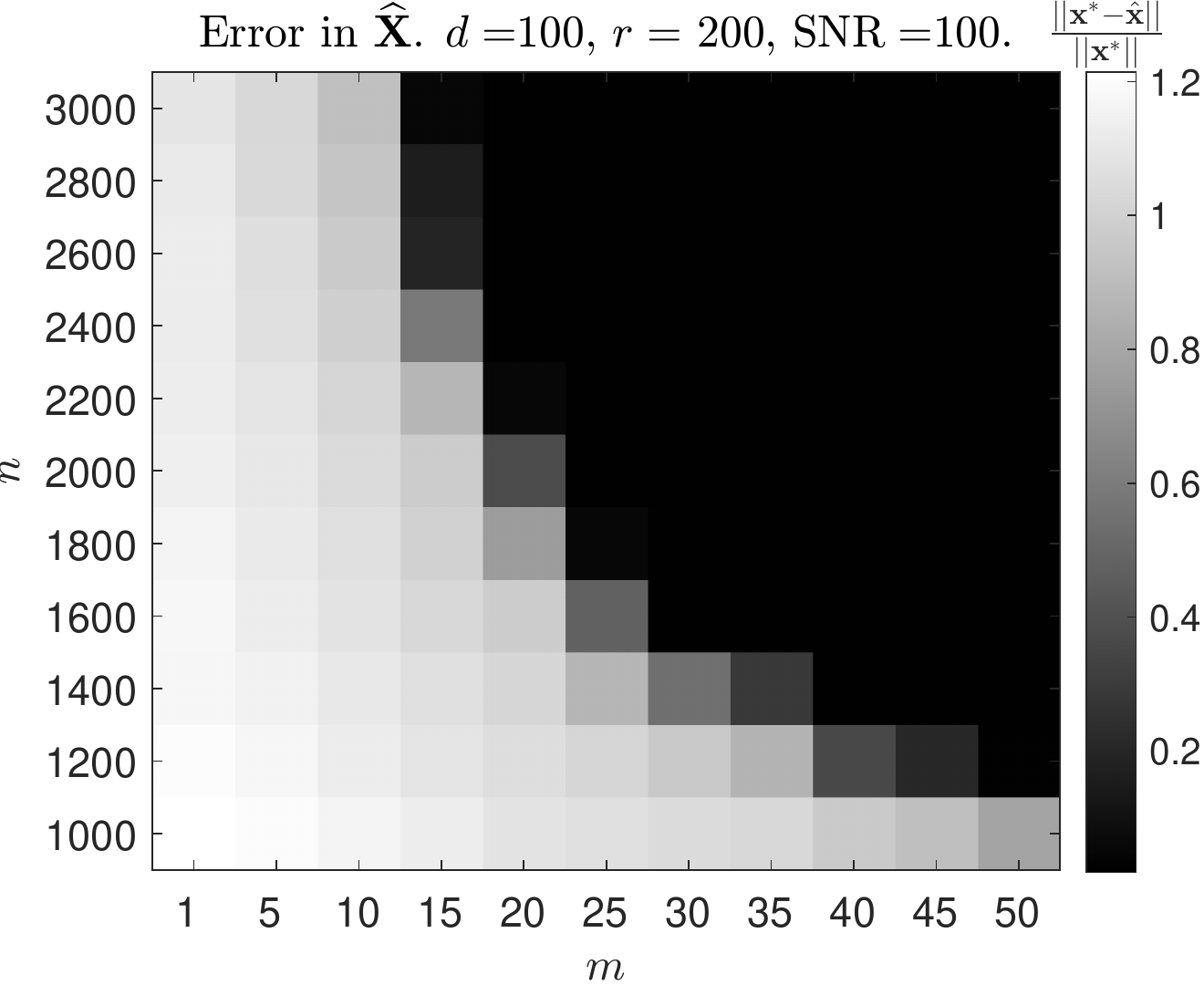}}
    \caption{}
    \label{fig:phase_transition_X}
    \end{center}
    \end{subfigure}%
     \begin{subfigure}[t]{0.23\textwidth}
    \begin{center}
    {\includegraphics[width = 0.99 \linewidth]{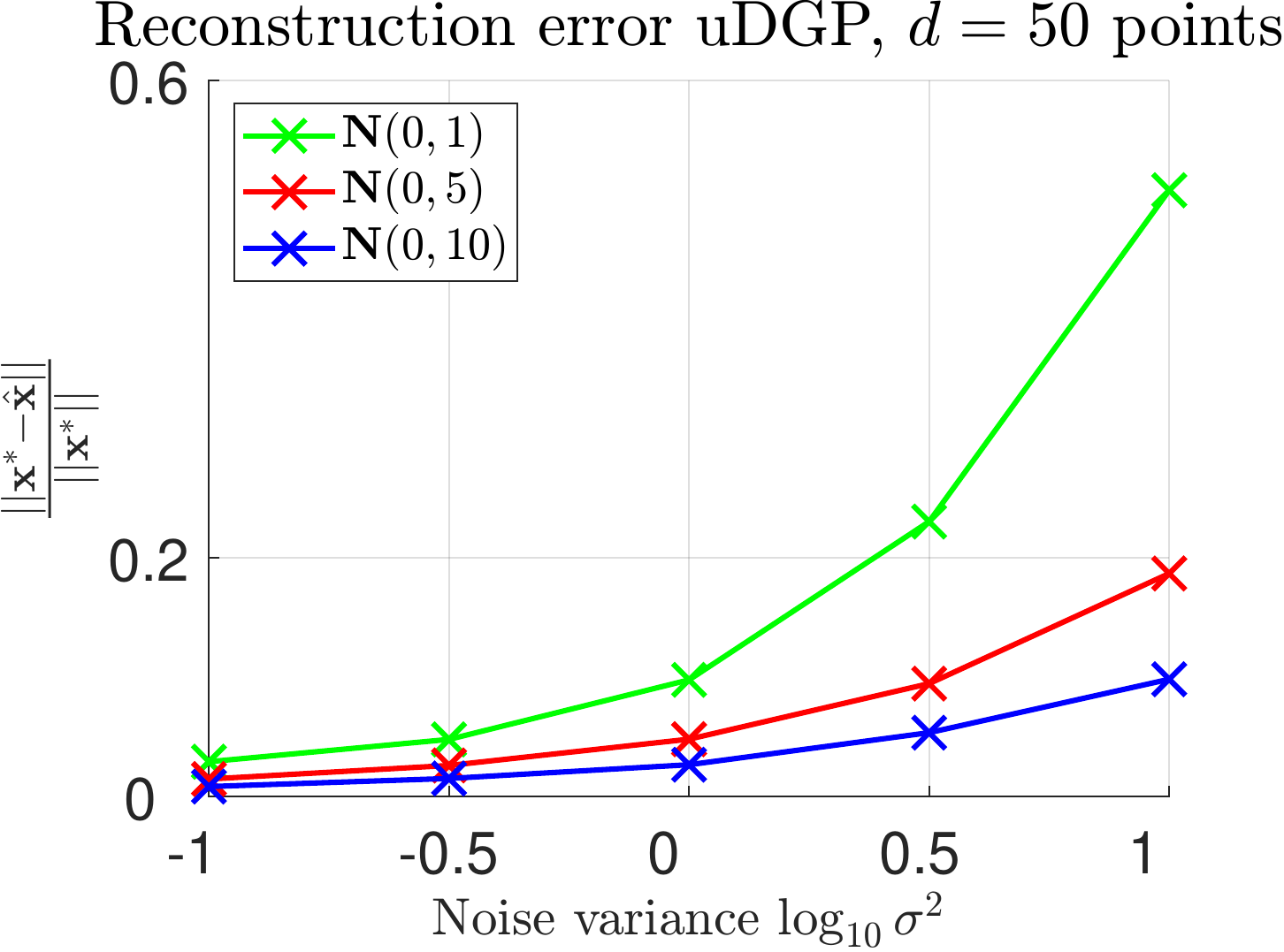}}
    \caption{}
    \label{fig:udgp}
    \end{center}
    \end{subfigure}%
    \caption{\textbf{Synthetic simulations} for $\Y = \P^*\B_{n \times d} \X^*_{d \times m} + \W$. (a). Figure plots the fractional Hamming distortion $d_{H}/n$ against block size $r$. The Hamming distortion  is the number of mismatches in estimate $\widehat{\P}$ of $\P^*$ and is defined as $d_H = \Sigma_{i}\mathbbm{1} (\widehat \P(i) \neq \P^*(i))$, where $\P(i)$ denotes the column index of the $1$ entry in the $i^{th}$ row of the permutation matrix $\P$. (b,c). Figures plot $d_H/n$ and the relative error $\lVert \widehat \X - \X^* \rVert/ \lVert \X^* \rVert$, respectively, against $(m,n)$ for fixed $(d,r)$. (d). \textbf{uDGP}. Figure plots the relative error in the estimated point coordinates against increasing noise variance $\sigma^2$. The $d=50$ points are sampled i.i.d. from the normal distribution with variances $\{1,5,10\}$, and the permutation $\P^*$ is $r$-local with blocks of sizes $(d-1,\cdots,2)$. }
    \label{fig:syn_sim}
    \vspace{-1.25em}
\end{figure*}
\begin{figure}[t!]
    \begin{subfigure}[t]{0.23\textwidth}
    \centering
    \includegraphics[width = 0.99\linewidth]{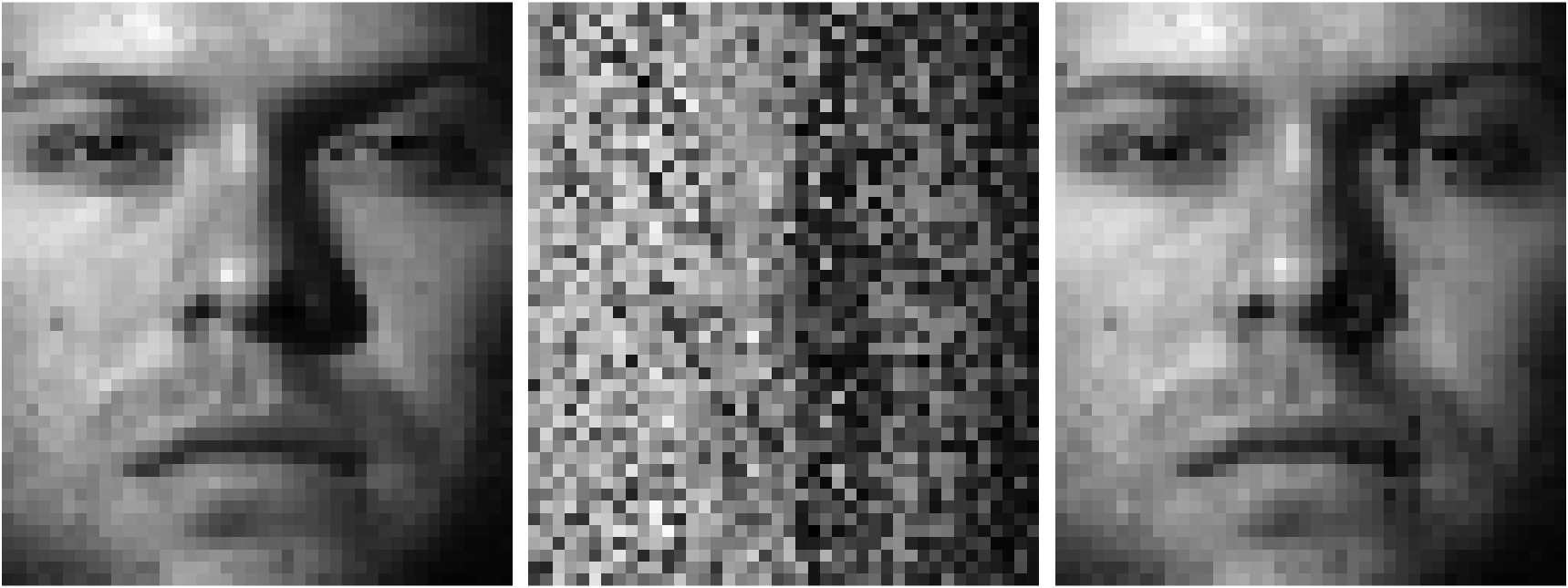}
    \caption{$\textrm{PSNR} = 27.04$ dB}
    \label{fig:big_r}
    \end{subfigure}%
    \begin{subfigure}[t]{0.23\textwidth}
    \centering
    \includegraphics[width = 0.99\linewidth]{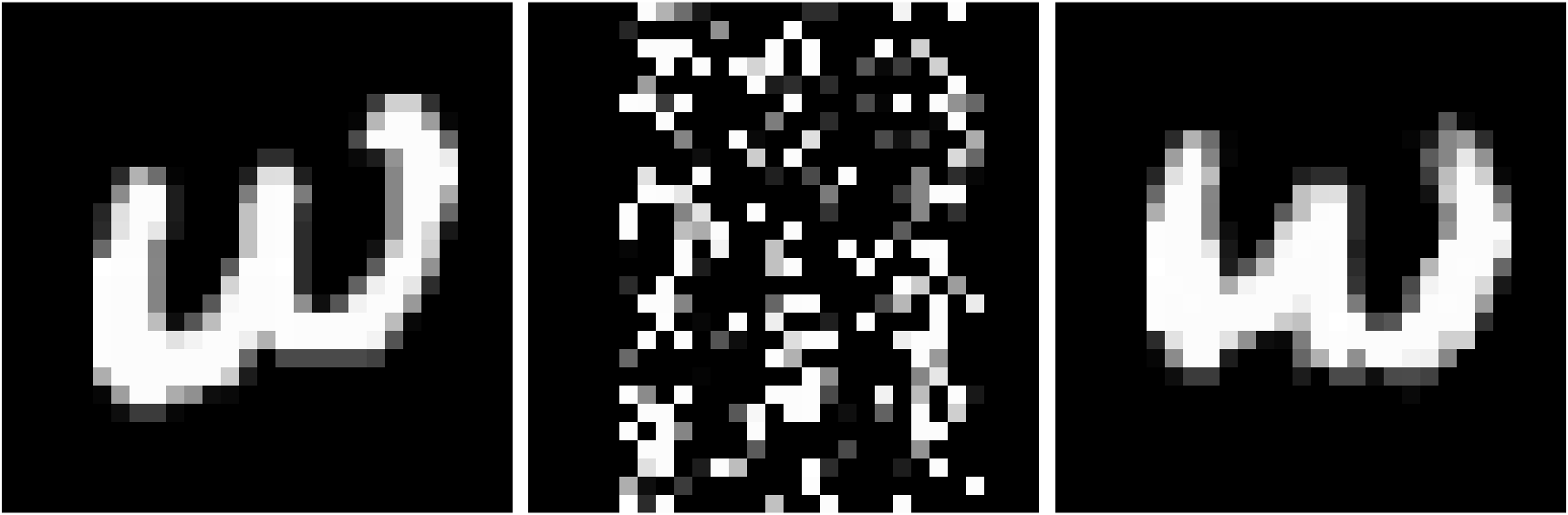}
    \caption{$\textrm{PSNR} = 11.79$ dB}
    \label{fig:small_r}
    \end{subfigure}    
    \vspace{-0.5em} \caption{(a).  Left. Unscrambled image $\y \in \real^n$, $n=48 \times 42$ from the YALE B dataset. Middle. Scrambled input image $\P^*\y$ via an $r$-local permutation, $r=96$. Right. Reconstructed image $\hat \y = \widehat{\P}^{\intercal}\P^*\y$.
    (b).  Left: Unscrambled image $\y \in \real^n$, $n=28 \times 28$ from the MNIST dataset.  Middle: Scrambled input image via an $r$-local permutation, $r=28$. Right: Reconstructed image. The peak signal to noise ratio $\textrm{PSNR}$ is defined in terms of the mean square reconstruction error $e = \frac{1}{n} \lVert \y -  \hat \y \rVert_2^2$ as $\textrm{PSNR} = 10 \log_{10}(\frac{1}{e^2})$. \vspace{-0.85em}}
    \label{fig:yale}
\end{figure}
\vspace{-1.35 em}\subsection{Synthetic simulations}
\vspace{-0.55 em}
\textbf{Data generation}: The entries of the signal matrix $\X^*$, the sensing matrix $\B $, and the noise $\W$ are sampled i.i.d. from the normal distribution. The matrix $\W$ is subsequently scaled by $\sigma$ to set a specific $\textrm{SNR}$. The permutation $\P^*$ is picked uniformly at random from the set of $r$-local permutations. All results are averages of $75$ Monte Carlo (MC) runs.

\textbf{Baselines}: We compare our algorithm to four algorithms proposed in \cite{snr,Levsort,zhang2019permutation,ojsp}. First is a biconvex relaxation of \eqref{eq:obj} proposed in  \cite{snr} and solved via the alternating direction method of multipliers (ADMM) algorithm.
Second is the Levsort algorithm in \cite{Levsort}. For $m=d$ views and $\sigma^{2} = 0$, the Levsort algorithm provably recovers $\P^*$ exactly. The third algorithm in \cite{zhang2019permutation} approximates $\widehat \X  \simeq   \B^{\intercal}\Y$ in \eqref{eq:P_update} and  estimates $\widehat \P = \argmin_{\P \in \Pi} \langle  \Y\Y^{\intercal}\B\B^{\intercal}, \P\rangle$. The approximation is justified when  matrix $\B$ is Gaussian i.i.d. and the rows of $\B$ are partially shuffled. In order to ensure a fair comparison with the algorithms in \cite{snr,Levsort,zhang2019permutation}, we specify an additional constraint to constrain the permutation solutions to be $r$-local.  For each MC run, the ADMM penalty parameter $\rho$ for \cite{snr}  was tuned in the range $10^{-3}$ to $10^{5}$ and the best results were retained. We also compare to the algorithm in \cite{ojsp} that solves a quadratic assignment problem to estimate $\P^*$. We refer to the algorithms in \cite{snr,Levsort,zhang2019permutation,ojsp} as `Biconvex', `Levsort, `One-step', and `RLUS', respectively.

\textbf{Results}: In Figure \ref{fig:benchmark}, we compare our proposed algorithm to the baselines. One-step  \cite{zhang2019permutation} fails to recover $\P^*$ for even small values of $r$, underscoring that algorithms to recover sparse permutations cannot be adapted to the $r$-local model. The Levsort algorithm \cite{Levsort} also fails to recover $\P^*$. The Biconvex  algorithm  $\cite{snr}$ only recovers $\P^*$ for $r = 20$.  RLUS \cite{ojsp} gives comparable recovery of $\P^*$ when $r \leq 50$, but is outperformed by the proposed algorithm at higher values of $r$. In Figures \ref{fig:phase_transition_P} and \ref{fig:phase_transition_X},  we set $(d,r) = (100,200)$ and vary $(n,m)$. The errors in $\widehat\X,\widehat \P$ decrease as $n$ increases. This is because the initialization to the algorithm (see section \ref{sec:algo}) improves with increasing $n$. The estimates also improve with increasing $m$. This observation agrees with the discussion in section \ref{sec:intro} and the result in \eqref{eq:inach}.
\vspace{-1.25em}\subsection{Scrambled image restoration} 
\vspace{-0.5em}We consider a variation of the problem  in \cite{gumbel_sinkhorn}. Therein, given scrambled images $\P^*\y$ and unscrambled training data $\y$, a neural network is trained to unscramble $\P^*\y$.  For our experiment, the images are drawn from the YALE B dataset \cite{Yale} and the MNIST dataset \cite{MNIST}.   For each class in the MNIST (YALE B) dataset, the matrix $\B \in \real^{n \times d}$ contains the $d=5$ $(10)$ principal components of the unscrambled data.  A class in the MNIST (YALE B) dataset comprises all images corresponding to a single digit (particular face). The problem is to recover the unknown permutation $\P^*$ given the scrambled image $\P^*\y$ and the matrix $\B$. The  setup is summarized in Figure \ref{fig:yale}.  We compare our algorithm to RLUS \cite{ojsp} and the results are given in Table $1$. The results show that the proposed algorithm outperforms RLUS \cite{ojsp}. 
\begin{table}
\begin{center}
  \centering
  \begin{tabular}{|c|c|c|}
    \cline{2-3}
    \multicolumn{1}{c|}{} & Proposed & RLUS \cite{ojsp}  \\ \hline
    MNIST & $\mathbf{12.31} \pm 2.57$    & $12.03 \pm 2.47$     \\ \hline
    YALE & $\mathbf{28.30} \pm 2.87$  & $26.43 \pm 2.95$     \\ \hline
  \end{tabular}
\end{center}
\vspace{-1em}\captionof{table}{Scrambled image restoration problem. For each dataset, the $\textrm{PSNR}$ values (dB) are averaged over $10$ classes.} \label{tab:img_psnr} 
\vspace{-1.05em}
\end{table}
\vspace{-0.55em}
\subsection{1-d uDGP}
\vspace{-0.55em}
We consider the $1$-dimensional uDGP problem (Section \ref{sec:udgp}) where the distances are corrupted with i.i.d. Gaussian noise of variance $\sigma^2$. The results in Figure \ref{fig:udgp} show that the relative error in the recovered points is low even for high noise variance. The results are noteworthy because the general $1$-d uDGP problem with noisy distances is NP-hard \cite{cieliebak2004measurement}.
\vspace{-0.85em}\section{Conclusion}
\vspace{-0.75em}
We propose a proximal alternating minimization algorithm for the unlabeled sensing problem and apply it to the setting where the unknown permutation is $r$-local. The resulting algorithm is efficient and theoretically converges to a first order stationary point. Experiments on synthetic and real data show that the algorithm outperforms competing baselines. We formulate the $1$-d unassigned distance geometry problem (uDGP) as an unlabeled sensing problem with a structured measurement matrix. Future work will explore information-theoretic inachieviability results for the $r$-local permutation model and uDGP. 
\clearpage
\bibliographystyle{IEEEbib}
\bibliography{strings}
\end{document}